\documentclass[12pt,preprint]{aastex}
\newcommand{\kms}{\mbox{km s$^{-1}$}}

\newcommand{\caii}{\ion{Ca}{2} H and K~}
\newcommand{\mv}{\ensuremath{M_{\mbox{\scriptsize V}}}}
\newcommand{\msun}{M$_{\odot}$}

\received{}
\accepted{}

\slugcomment{}

\shortauthors{Rauscher \& Marcy}
\shorttitle{Ca II H and K Lines in M Dwarfs}

\begin{document}

\title{\caii Chromospheric Emission Lines \\ in Late K and M Dwarfs$^{1}$}
\author{Emily Rauscher\altaffilmark{2},
Geoffrey W. Marcy\altaffilmark{2}}

\altaffiltext{1}{Based on observations obtained at the W. M. Keck
Observatory, which is operated jointly by the University of California
and the California Institute of Technology.  Keck time has been
granted by both NASA and the University of California.}

\altaffiltext{2}{Department of Astronomy, University of California,
  Berkeley, CA USA 94720}

\email{gmarcy@astro.berkeley.edu}

\begin{abstract}  
We have measured the profiles of the \caii chromospheric emission
lines in 147 main sequence stars of spectral type M5-K7 (masses
0.30-0.55 \msun) using multiple high resolution spectra obtained
during six years with the HIRES spectrometer on the Keck 1 telescope.
Remarkably, the average FWHM, equivalent widths, and line luminosities
of \caii {\it increase} by a factor of 3 with increasing stellar mass
over this small range of stellar masses.  We fit the \caii lines with
a double Gaussian model to represent both the chromospheric emission
and the non-LTE central absorption.  Most of the sample stars display
a central absorption that is typically redshifted by
$\sim$0.1 \kms~relative to the emission.  This implies that the
higher-level, lower density,  chromospheric material has a smaller outward
velocity (or higher inward velocity) by 0.1 \kms~than the lower-level material in
the chromosphere, but the nature of this velocity gradient remains
unknown.  The FWHM of the \caii emission lines increase 
with stellar luminosity,
reminiscent of the Wilson-Bappu effect in FGK-type stars.  Both the
equivalent widths and FWHM exhibit modest temporal variability in
individual stars. At a given value of \mv, stars exhibit a spread in
both the equivalent width and FWHM of \caii, due both to a spread in
fundamental stellar parameters including rotation rate, age, and
possibly metallicity, and to the spread in stellar mass at a given
\mv.  The K line is consistently wider than the H line, as expected,
and its central absorption is more redshifted, indicating that the H
and K lines form at slightly different heights in the chromosphere
where the velocities are slightly different.  The equivalent width of
H$\alpha$ correlates with \caii only for stars having \ion{Ca}{2}
equivalent widths above $\sim$2 \AA, suggesting the existence of a
magnetic threshold above which the lower and upper chromosphere become
thermally coupled.
\end{abstract}


\section{Introduction}

M dwarfs have received considerable attention regarding their internal
stellar structure, their atmospheres, and their distribution in the
Galaxy (see reviews by \citet{Liebert1994}, \citet{Burrows1993} and
\citet{Bochanski2005} and references therein).  However, their atmospheric
structure, dynamics, and magnetic processes
remain poorly understood.  M dwarfs show clear evidence of
chromospheric heating, as revealed by emission lines in optical and
UV wavelengths \citep{Giampapa1989, Gizis2002, Reid2002} 
that constrain models of chromospheric structure \citep{Cram1987}.
They also contain hot coronae, as detected by their X-ray emission,
with $L_X/L_{bol}$ up to $10^{-3}$ \citep{Caillault1996}.
\citet{Robrade2005} and others have analyzed X-ray flaring in M
dwarfs, and many observations of flaring activity have been made at
other wavelengths \citep{Eason1992, Hawley2003}.  Although
hot outer atmospheres and flaring activity have been studied in stars of
F, G, and K spectral types, M dwarfs deviate from chromospheric/coronal
relations found in those stars, indicating a qualitative difference
in atmospheric structure, velocity fields, and magnetic morphology
\citep{Schrijver1987, Rutten1989, Valenti2003}.

Observations of chromospheres and coronae in M dwarfs have led to
models that include significant amounts of nonradiative, magnetic
heating in their outer atmospheres \citep{Giampapa1982,
Johnskrull1996}.  The observed scatter in X-ray luminosities and
chromospheric emission line fluxes can be attributed to temporally
changing surface inhomogeneities \citep{Panagi1993} or differences in
filling factor between stars \citep{Reid1995}.  \citet{Mullan1996}
showed that acoustic dissipation may provide enough coronal heating
for stars dominated by quiet regions, but magnetic heating is required
to account for the coronal heating in the more active M dwarfs that
likely have sizable covering factors of kilogauss magnetic fields
\citep{Johnskrull1996, Valenti2003}.

Here, we probe the atmospheres of M dwarfs with near-UV spectra at the
resonant ``H and K'' transitions of \ion{Ca}{2} at 3968 and 3933 \AA~respectively.  We use sufficiently high resolution to resolve
structure in the emission line profiles and to extract line widths,
asymmetries, and central absorptions.  We examine a sample of
147 late K and M dwarfs from spectra previously obtained by the
California and Carnegie Planet Search.  Multiple spectra for each star
yield mean values over time and any temporal changes.  We compare the
\caii lines to H$\alpha$ equivalent widths, providing a tie to coronal
gas.

In Section~\ref{sec: observations} we describe the stellar sample and
the Keck spectra.  In Section~\ref{sec: analysis} we describe the
double Gaussian model of the lines.  In Section~\ref{sec: results} we
present the resulting observed line parameters, and in
Section~\ref{sec: discussion} we offer discussion and summary.

\section{Observations} \label{sec: observations}

All spectra used in this study were taken from the California and
Carnegie Planet Search Program, with a set-up described by
\citet{Wright2004} and briefly summarized here.  From 1997-2004, the
Planet Search Program has been monitoring the velocities of 700 FGKM
stars with the goal of detecting extrasolar planets, as described by
\citet{Butler1996}.  The spectra were all obtained with the HIRES
spectrometer at Keck Observatory \citep{Vogt94}.  They span the
wavelength region, 3850 to 6200 \AA~with a resolution of $R$=55,000
corresponding to a spectrometer PSF with FWHM = 5.5 \kms~or 2.8
pixels (24 $\mu$ each) on the CCD.  Exposure times of typically 8
minutes yield a signal-to-noise ratio of 60 per pixel in the continuum
around the Ca II H and K lines for the majority of the M dwarfs, most
having V=9-11 studied here.  But the quality can be higher or lower
depending on seeing and B-V of the star.  The spectra are not
flux-calibrated.

The stars included here are those from the Planet Search Program
having B-V $>$ 1.3 corresponding to a spectral type of about K7.  All
colors were taken from the Hipparcos catalogue \citep{ESA97}.  They
were originally selected for that search by sieving the Gliese Catalog
and Hipparcos catalog for late K-type and M dwarfs that were both
brighter than V = 11.5 and devoid of known stellar companions within 2
arcsec, to avoid contamination at the spectrometer slit.  The stars
were chosen without regard to previously known emission lines in their
spectra.  Indeed, most such stars exhibit emission at \caii (but not
necessarily at H$\alpha$).  Thus, this present sample is
representative of the single, K7-M5 dwarfs within $\sim$15 pc, though
some are slightly farther away.\footnote{Other exceptions stem from
slight errors in photometry, notably HD~90875 that had a B-V color
originally mislabeled as 1.70.  It is actually spectral type K5.}  The
limit at spectral type M5 stems largely from their faintness for the
purposes of high resolution spectroscopy, most such stars 
being fainter than V = 11.5.

\begin{figure}[ht!]
\begin{center}
\includegraphics[angle=90,width=0.75\textwidth]{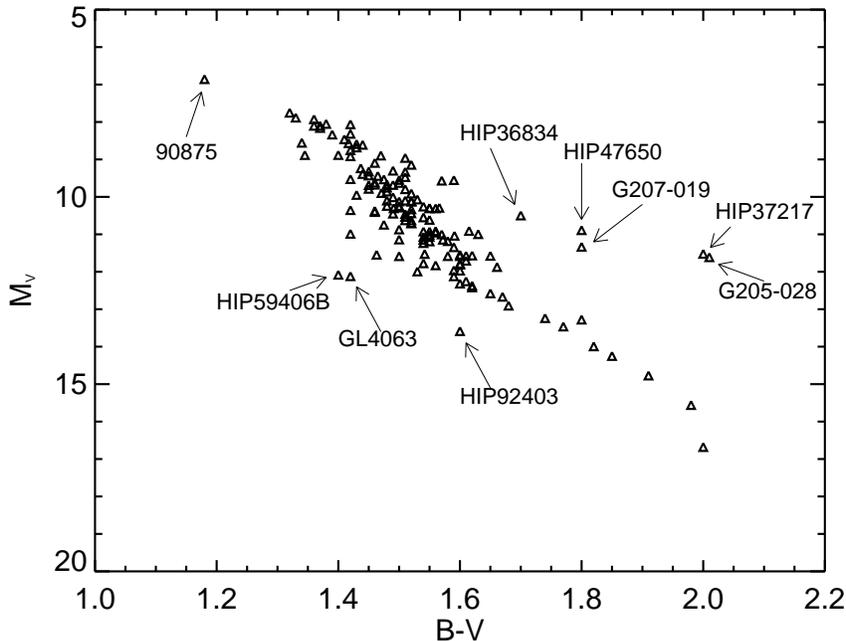}
\caption{The color-magnitude diagram for the 147 K7-M5 dwarfs in the
  stellar sample.} \label{fig: hr_diagram}
\end{center}
\end{figure}

The resulting sample of 147 stars is shown plotted in a
color-magnitude diagram in Figure~\ref{fig: hr_diagram}.  The stars
departing farthest from the main sequence are labeled, with the departures 
due most likely to stellar companions, extraordinary metallicity, or simply to
poor photometry.  The width of the main sequence in Figure~\ref{fig:
hr_diagram} indicates the limited value of B-V as a diagnostic for
$T_{\rm eff}$ for M dwarfs, due to its sensitivity to metallicity
caused by line blanketing \citep{Gizis1997}.  Table 1 lists for each
star its common names, spectral type, \mv, B-V, the number of
observations, and the time range over which it was observed.

We obtained a total of 2408 spectra for the 147 stars.  Each spectrum
was reduced as part of the planet-search automated pipeline.  The raw
CCD images were biased-subtracted, the flat-fielding was done with a
quartz lamp through the same spectrometer setup, the spectral order
locations were mapped and fitted with a third order polynomial, and
the final one-dimensional spectra were extracted by summing along the
column within each order.  These spectra give the number of photons in
each pixel of width 2 \kms.

\section{Analysis} \label{sec: analysis}

Inspection by eye of the \caii emission line profiles from the 147
late-K and M dwarfs reveals considerable structure. The widths of the
\caii emission lines in the K7-M0 dwarfs are clearly wider than those
of the M1-M5 dwarfs.  Many emission lines are double-peaked,
reminiscent of the central ``absorption'' reversal well known to exist
in \caii profiles of G and K dwarfs.  Many M dwarfs that do not
exhibit double peaks appear to have flat-topped emission lines, as if
weak central absorption were suppressing the peak of the emission.
Also visible to the eye are asymmetries in the double-peaked emission
profiles, with the central absorption being shifted redward relative
to the emission.  The blueward peak is almost always taller than the
redward one, providing a sensitive measure of relative Doppler shift
of the emission and central absorption.  The \caii profiles stem from
the non-LTE line transfer and velocity fields in the low density
chromospheres of M dwarfs. The central absorption stems from the
dependence of the source function on height in the optically thick
chromosphere rather than from actual absorption.  The variety of
visible structure in the line profiles calls for quantitative
measurements, in advance of future modeling of the line transfer.

Our analysis of the \caii lines was performed by fitting them with the
sum of two Gaussian curves.  We tried other functional forms, such as
flat topped functions, but the double-Gaussian yielded the smallest
value of $\chi^2$ and (remarkably) reproduced all off the observed
structure in the line profiles.  The model consisted of a positive
Gaussian representing the chromospheric emission that emerges from the
broad bottoms of the photospheric Ca II absorption features.  Added to
that is a second Gaussian of negative amplitude representing non-LTE
``absorption'' at line center.  A constant background level was left
as a final parameter, representing the base of the photospheric
absorption.  The height, width, and central position of both Gaussians
were free to vary as well as the background level, for a total of 7
parameters used to minimize $\chi^2$.  This model provides excellent
fits to those emission lines that exhibit either obvious central
absorption or flat tops.  For those emission lines without an obvious
central absorption, we found that the double Gaussian commonly fit the
line profiles well, especially at the top, better than did a single
Gaussian.

\begin{figure}[ht!]
\begin{center}
\includegraphics[angle=90,width=0.6\textwidth,height=.25\textheight]{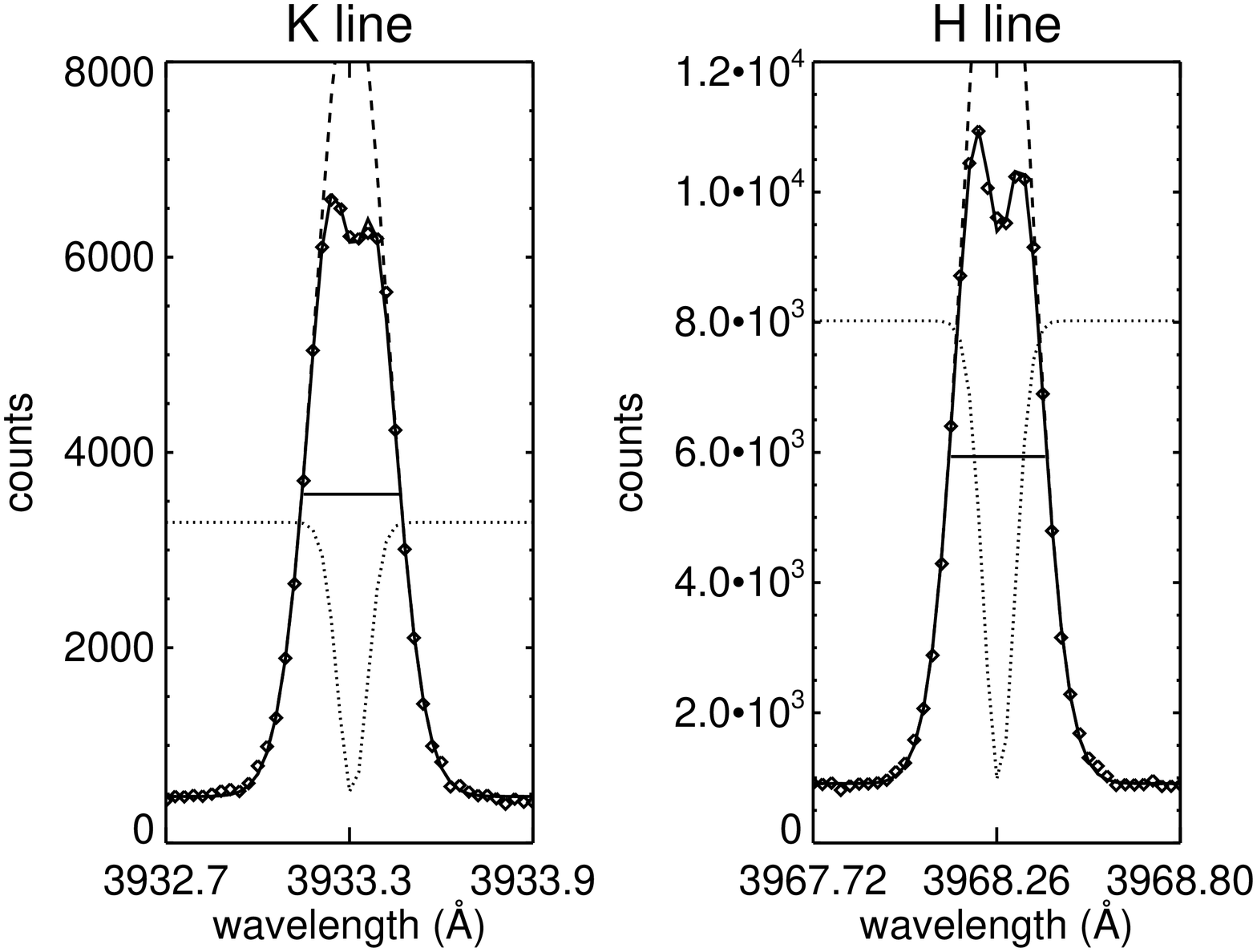}
\includegraphics[angle=90,width=0.6\textwidth,height=.25\textheight]{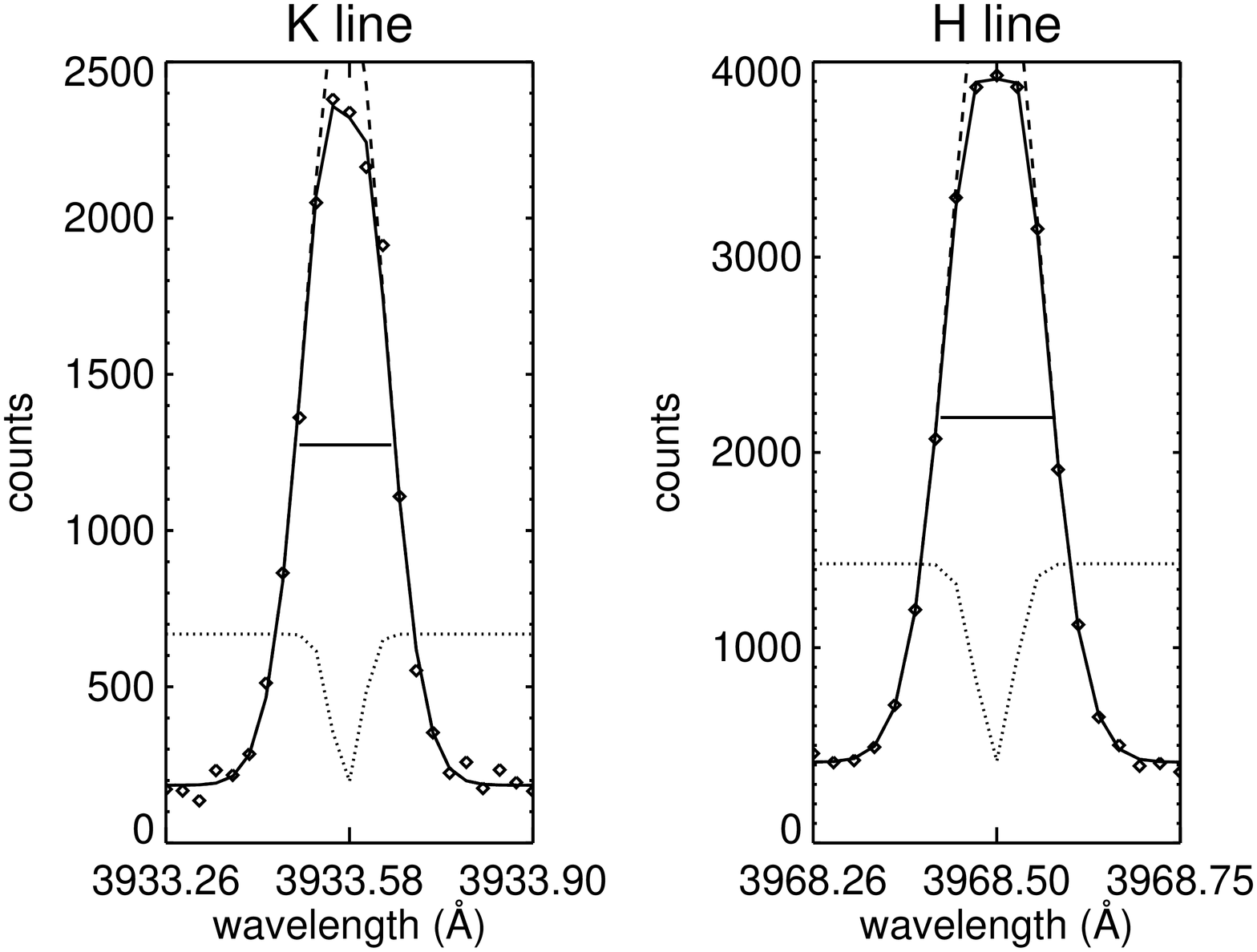}
\includegraphics[angle=90,width=0.6\textwidth,height=.25\textheight]{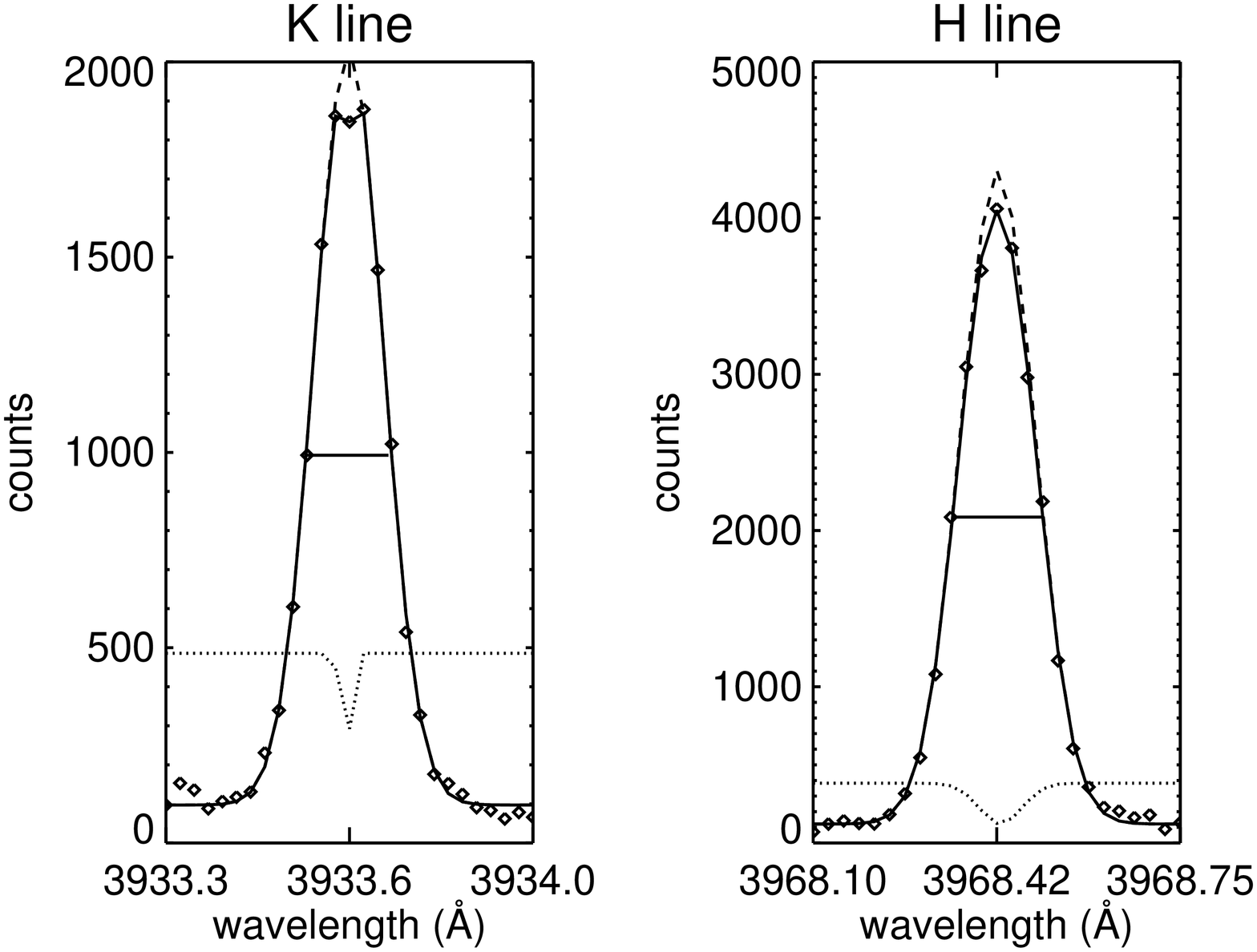}
\caption{Representative observed \caii profiles (dots) and the
double-Gaussian model fits (solid line) to them, for three stars.
The two separate Gaussians (emission and absorption) are plotted as
dashed and dotted lines, respectively.  Note that for flat-topped
lines, the negative Gaussian serves to produce an adequate fit. 
The FWHMs are shown for each line as a horizontal line at the mid-height.
Top spectrum is of GJ 809, middle spectrum of GJ 251, and bottom spectrum of GJ 406.} 
\label{fig: example_fit}
\end{center}
\end{figure}

Figure~\ref{fig: example_fit} shows our model fit to the H and K lines
from spectra of three stars, illustrating different line shapes:
obviously double-peaked lines, flat-topped lines, and thin lines.  In
each case the emission Gaussian alone fits the edges of the line but
overshoots the top, which can then be accommodated by the second
Gaussian of negative amplitude.

From these fits we derived a set of parameters for each line including
the equivalent width (EW), the full width at half maximum (FWHM), and
the wavelength difference between the centers of the two Gaussians
($\Delta \lambda_c$), indicating a relative Doppler shift between the
emission and absorption.  For those profiles that were clearly
double-peaked, we determined the ratio of the maximum counts in the
blue peak to the maximum counts in the red peak (V/R).  The equivalent
width was determined by dividing the area in the line by the continuum
around the line.  The area under the line was easy to measure.  We
subtracted the constant background found by our fit from the line
and integrated the counts within the line.

\begin{figure}[ht!]
\begin{center}
\includegraphics[angle=90,width=0.75\textwidth]{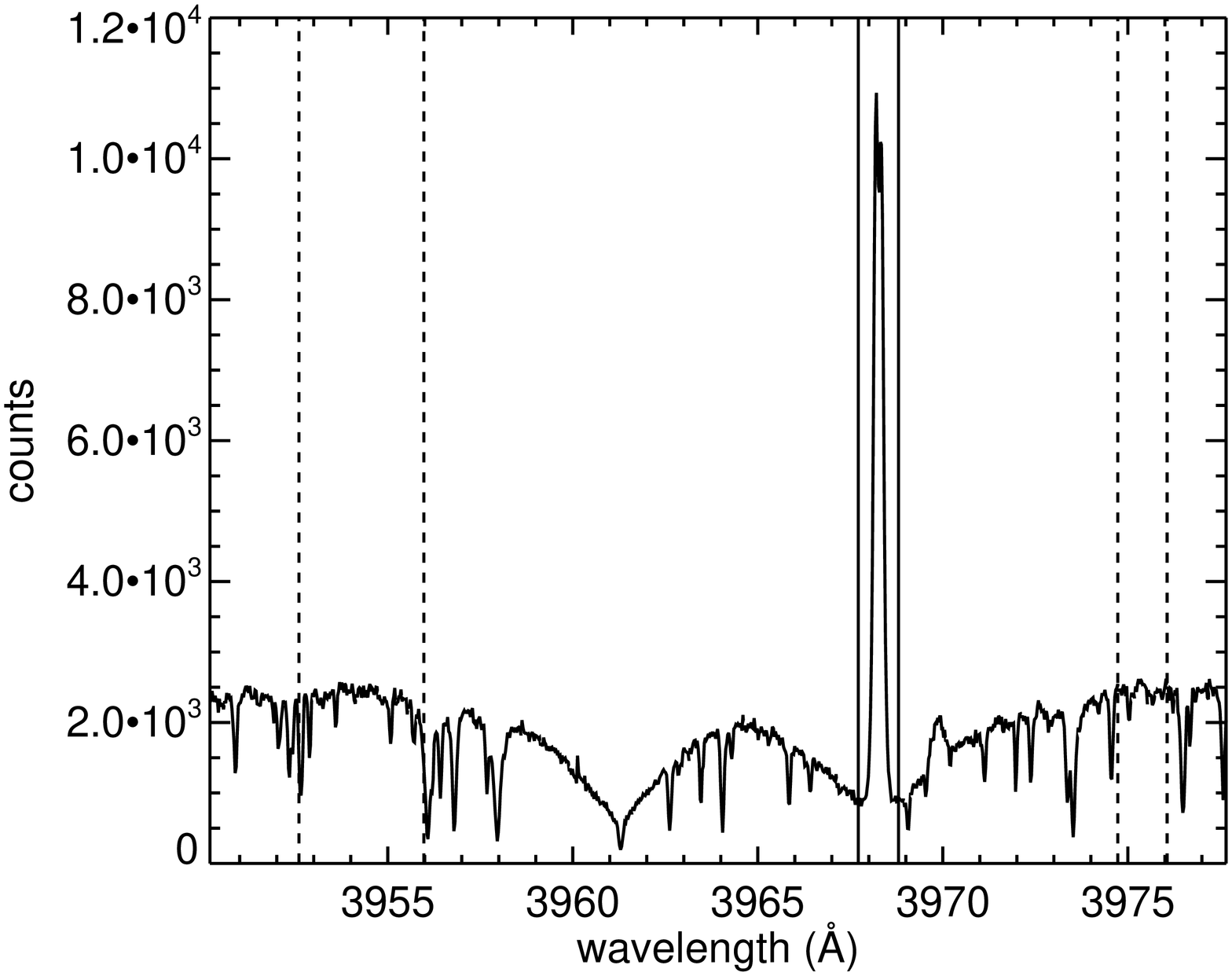}
\caption{A representative spectrum (from GJ 809), showing the \ion{Ca}{2} H 
line emission and the continuum in its vicinity.  The two vertical solid lines mark 
the wavelength region used for fitting the emission line.  The two
pairs of vertical dashed lines mark the wavelength ranges used to measure an average
continuum value.  Peculiar to this spectrum is the visible presence of
H$\epsilon$ emission at 3970 \AA, just redward of the H line.} 
\label{fig: example_cont}
\end{center}
\end{figure}

The continuum determination presented challenges.  The spectra of M
dwarfs in this region contain many absorption lines, leaving no
obvious choice for a constant value to assign to the continuum.  We
determined a continuum value by averaging the number of counts within
two wavelength bins located on either side of the broad photospheric H
line, as shown in Figure~\ref{fig: example_cont}.  These continuum
regions are analogous to those used in the determination of the
well-known Mt. Wilson $S$ values. (The locations of these ranges were
set with reference to the center of the emission line so as to be
constant within the rest frame of each star.)  We performed a spline
on the double-Gaussian fitted curve to find the width at half of the
background-subtracted maximum.

We used $\Delta \lambda_c$ and V/R as measures of the asymmetry of the
lines.  We defined $\Delta \lambda_c$ as the center of the absorption
Gaussian in our model minus the center of the emission Gaussian, so
that a value greater than zero corresponds to a redshifted absorption
relative to the emission.  As previously used by \citet{Smith2004},
V/R is only applicable for double-peaked lines, i.e. those with a
visible absorption at line center.  For these lines, we found the ratio of
the maximum number of counts in the blue side of the emission line to
that on the red side.  Lines having a blue peak higher than the red
peak (as typically seen here) have V/R $>$ 1.  Segregation of red and
blue peaks was formally made (without operator intervention) using the
location of the center of the emission Gaussian.  The parameter
$\Delta \lambda_c$ is applicable to more stars than was V/R since it
can be calculated for any line, whereas V/R requires a double-peaked
line.  Only 40\% of the stars here were double-peaked.

Our automated fitting procedure occasionally failed to find good fits
to the H and K lines in our 2408 spectra.  The process obtained
good fits for both of the lines for 63\% of the spectra and was not
able to find a good fit for either line in 15\% of the cases.  In many
cases the poor fits were simply due to spectra of such low
signal-to-noise ratio that the $\chi^2$ minimization algorithm failed.
Since we had multiple observations of each star, we ignored the
poorest fits from our subsequent analysis.

Four stars had no adequate fit of either line, with $\chi^2$ well
above unity.  Two of them, L 758-107\footnote{This star is internally
referred to as HIP 59406B.} and GL 745B, have emission lines that
were too weak to fit with any integrity.  The other two, GL 388 and
HIP 112460, have very strong H and K lines, but the reduced $\chi^2$
values were above 10 due to their distinctly non-Gaussian profiles,
and therefore we consider these fits dubious.  Of the 147 original
stars, 136 have at least one good measurement of both the H and the K
lines, and another 7 stars have a good quality measurement of at least one line.

\section{Results} \label{sec: results}

From each spectrum of the H and K lines for each star, we compiled 
the mean values of the best-fit parameters and their 
standard deviations.  The latter is a measure
of internal precision and temporal variability, to be discussed in
each subsequent section.  Tables 2 and 3
list the final measurements for the H and K lines, respectively, for
each star.  Columns 1-3 list the common names of each star, and column 4
gives the number of spectra analyzed with good fits.  Columns 5 and 6 give the
equivalent width and its standard deviation.  Columns 7 and 8 give the
FWHM and its standard deviation.  Columns 9 and 10 give $\Delta
\lambda_c$ and its standard deviation and columns 11 and 12 give the
mean and standard deviations for $V/R$.

Ideally, one would like to measure chromospheric parameters as a function
of stellar mass.    Unfortunately, we do not have secure photometric 
calibration of stellar mass for M dwarfs that would be immune to 
variations in chromospheric activity.  Nor do we have a mass calibration
immune to the unknown metallicities of the stars. 
Thus B-V and \mv~remain as modest proxies of stellar mass here,
with due concern about their integrity as such.
We proceed to examine how the measured chromospheric parameters 
depend on \mv~which is likely a better proxy for mass than B-V.
Clearly there is a great need for accurate masses for M dwarfs.

\subsection{Equivalent Widths and Line Widths of Ca II H and K} \label{sec: sptype}

Previous studies of M dwarfs have reported measurements of either line fluxes
(\citet{Robinson1990}, \citet{Panagi1993}) or equivalent widths as the
diagnostics stellar activity, most often using unresolved lines.
As our spectra are not flux calibrated, we report the equivalent
width of the \caii lines, obtained from the double-Gaussian fit.  
Such measurements are explicitly dependent on the
color, and hence \mv~and mass, of the star.  The nearby UV continuum decreases
with increasing color and lower mass.  Thus for a fixed luminosity in
the lines, the \caii equivalent widths would increase with increasing
stellar color, simply due to the decreasing continuum.

\begin{figure}[ht!]
\begin{center}
\includegraphics[angle=90,width=0.65\textwidth]{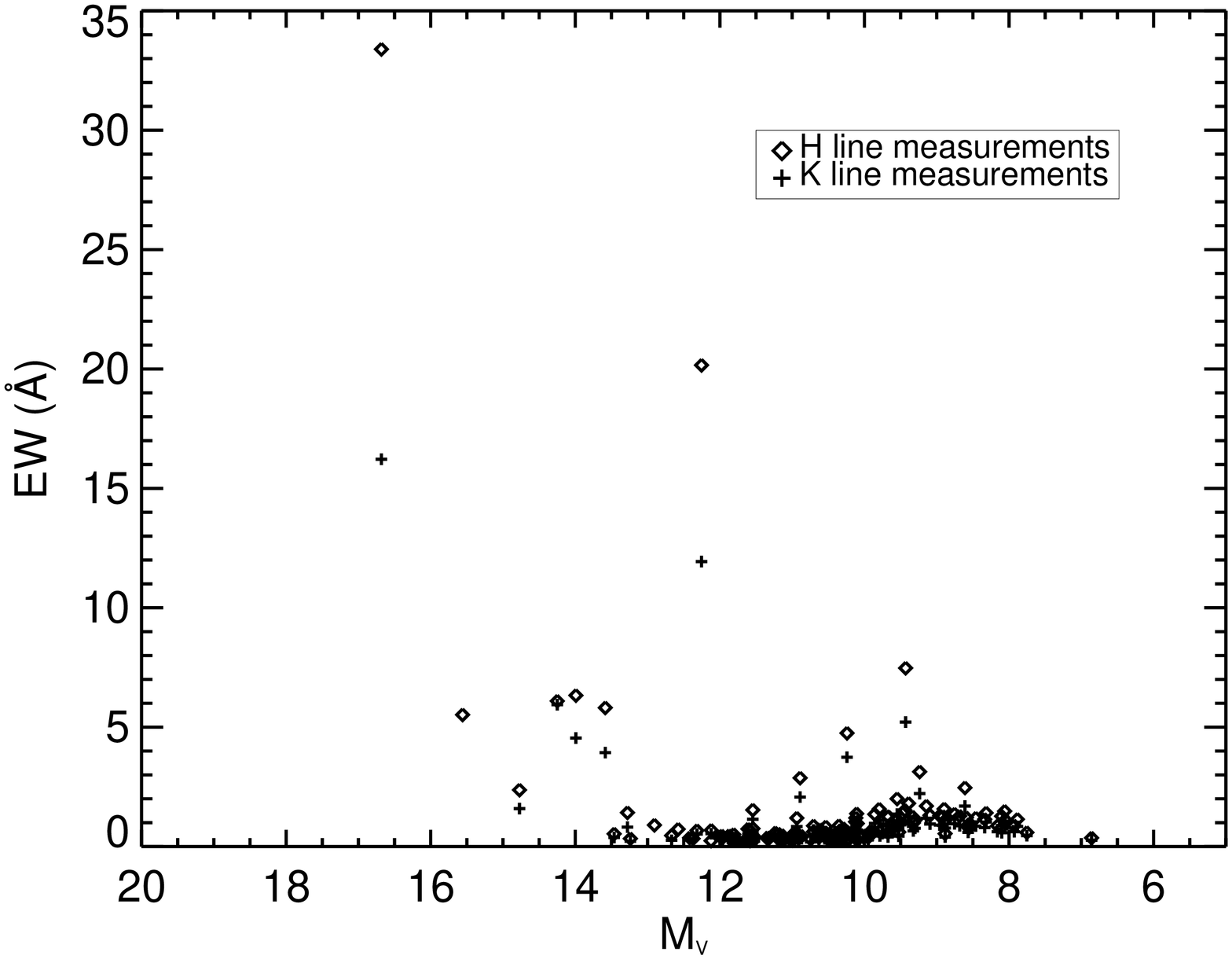}
\includegraphics[angle=90,width=0.65\textwidth]{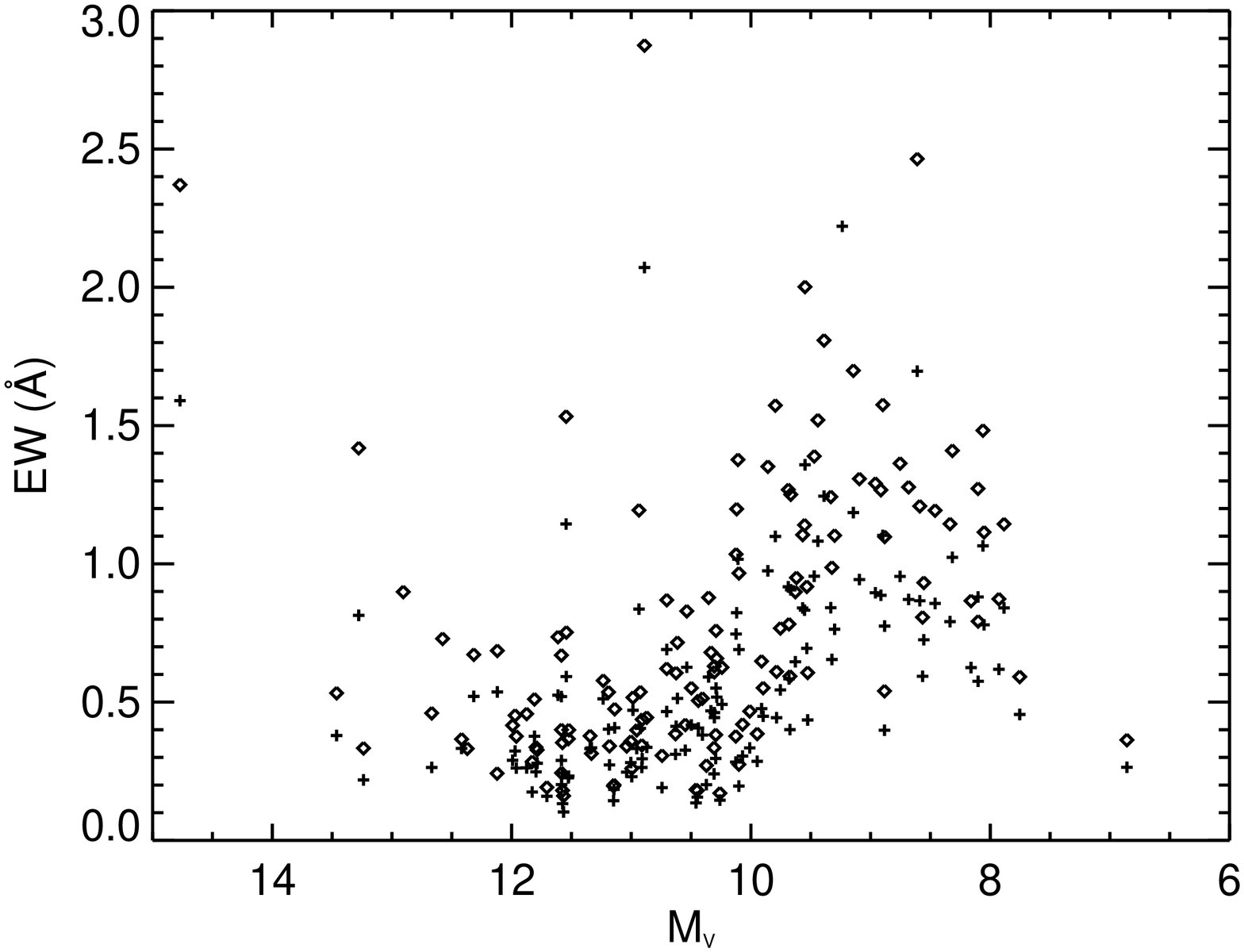}
\caption{The mean EW of \caii for each star versus its $M_V$.  Top:
  All stars.  Bottom: a close-up of the majority of stars that have EW
  $<$ 3 \AA.  The errors of $\sim$0.025 \AA~are less than the 
size of the symbols.} 
\label{fig: ew_sptype}
\end{center}
\end{figure}

Figure~\ref{fig: ew_sptype} shows the measured values of EW for the
\caii lines (diamonds and crosses, respectively) versus \mv.  The
figure shows that the EWs of \caii for late K and early M dwarfs (\mv
= 7-9) are typically $\sim$1 \AA.  However, the later type M dwarfs
(\mv = 9-13) have smaller EWs of \caii , typically $\sim$0.3 \AA.  Thus, the EW
of \caii increases by a factor of 3 toward decreasing \mv, 
and thus toward higher stellar mass.
The increase in EW with stellar mass is in the opposite direction 
from that which would be caused simply by the increasing 
UV continuum flux with stellar mass.  
{\it Thus, the luminosity in the
\caii lines increases, by at least a factor of 3, with
increasing mass from M5 to K7.}

Of some interest is the transition from partially to fully convective
stars that is expected to occur near 0.3 $M_\sun$ \citep{Burrows1997}.
Indeed, \citet{Gizis2002} found an increased level of activity (as
measured by H$\alpha$ EWs) at M2.5/3, which corresponds roughly to
this mass.  \cite{Mathioudakis1991} and \citet{Fleming1995} note that
0.3 \msun (where full convection sets in) may correspond to spectral
types M4-M5.  It is tempting to associate the remarkable factor of 3
rise in the chromospheric H and K lines seen here with a dynamo
explanation. Such an explanation may be appropriate.  Note, however,
that \citet{Gizis2002} found an increase in H$\alpha$ EW for later
type stars while here we find a decrease in the \caii emission for
those stars.  Nonetheless, Figure~\ref{fig: ew_sptype} does not reveal
a sharp discontinuity in EW at any \mv. Thus we find no convincing
evidence of a discontinuous change in the dynamo in the domain K7-M5.

Since the plotted values of EW are the means of multiple measurements
for each star, we can compare the vertical spread of the points to the
standard deviations seen for each star individually.  The standard
deviations of EW (listed in Tables 2 and 3) are typically $\sim$0.1
\AA~while the spread in EWs at any given $M_V$ is much larger.  Thus
it appears that the spread in EW at a given \mv~is greater than can be
accounted for by temporal variability.  The observed spread may be
caused by the degeneracy in mass at a given $M_V$; however, we suspect that
this degeneracy alone is not adequate to account for all of the
spread.  Rather, the chromospheric activity of a star is likely
influenced by other stellar parameters, notably rotation, that vary
from star to star at a given mass.  Indeed similar analysis in H$\alpha$
emission has shown that activity is related to both age and rotation
\citep{Stauffer1997}, which is likely the cause of variance of
\ion{Ca}{2} emission as well.

The FWHM of the \caii emission lines is plotted in Figure~\ref{fig:
width_sptype} versus \mv.  This plot shows a tight relation between
\caii line width and \mv, such that FWHM increases with brighter
\mv~and hence with stellar mass.  This is reminiscent of the
Wilson-Bappu effect in FGK stars as discussed in \citet{Giampapa1994}
and modeled by \citet{Ayres1979}, where the logarithm of the width of
chromospheric emission lines is observed to increase linearly with
decreasing absolute magnitude.  This remarkably tight relationship
between line width and \mv~for M dwarfs is not understood, to our
knowledge.  The conventional Wilson Bappu effect in FGK stars
represents a correlation between emission line width with luminosity
which in turn is directly related to evolutionary state (luminosity
class) and hence surface gravity.  In contrast, the M dwarfs all
reside nearly on the main sequence, none having evolved at all, albeit
having different masses, radii, and gravities.  Nonetheless, it is
apparent that the \caii line width in M dwarfs could similarly serve
to determine distances.  Ultimately, a line-transfer explanation of
the Wilson-Bappu effect in M dwarfs must include the variation of
chromospheric pressure and mass column as a function of stellar mass.

\begin{figure}[ht!]
\begin{center}
\includegraphics[angle=90,width=0.75\textwidth]{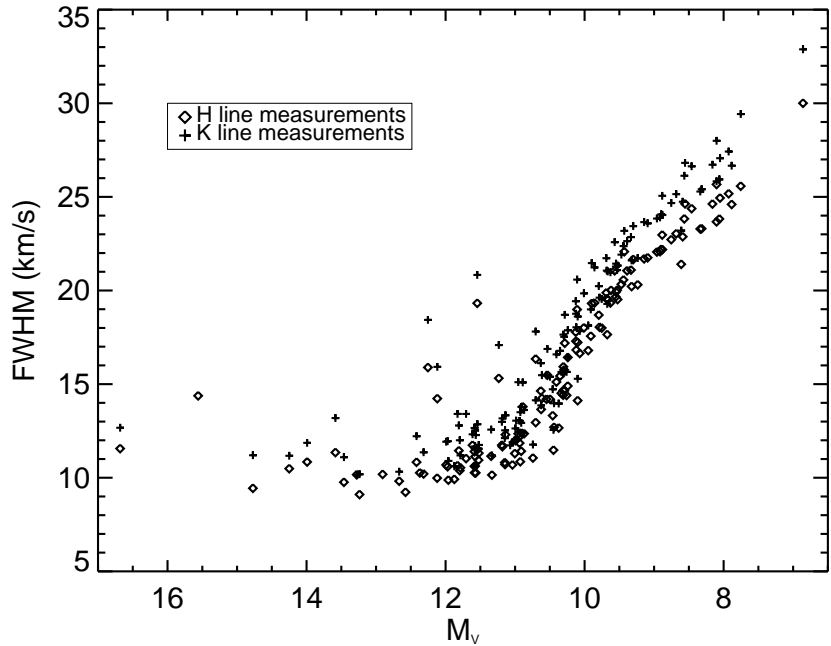}
\caption{The mean FWHM of the H and K lines for each star as a
function of \mv.  The FWHM increases with brighter \mv, in a
remarkably tight correlation.  The FWHM increases by a factor of
$\sim$3 from over this narrow range in spectral types, M5-K7. 
This correlation is  reminiscent of the Wilson Bappu effect known for
FGK stars.   The errors in FWHM are $\sim$0.3 \kms, smaller than the
size of the symbols.} 
\label{fig: width_sptype}
\end{center}
\end{figure}

Slight inflections are visible in Figure~\ref{fig: width_sptype}, but
the limitations discussed earlier associated with the use of $M_V$ as
a proxy for stellar mass may be the cause. 

Figure~\ref{fig: width_sptype} exhibits a lower limit of 9 \kms~for
the observed FWHM of the \caii lines.  This limiting width represents
the convolution of the lowest intrinsic line widths and the width of
the PSF of the spectrometer, 5.5 \kms.  Thus, considering a quadrature
sum of widths, the intrinsic line widths must have a lower limit of
7.1 \kms~for K7-M5 dwarfs.  There appear to be few, if any, stars from K7-M5
having lines narrower than 7.1 \kms, even among those fainter than \mv = 14.
Such a lower limit to the \caii line widths would indicate a
departure from the monotonic Wilson-Bappu relation set by the brighter
M dwarfs.  Perhaps some other broadening mechanism besides line transfer
effects sets a floor on the FWHM of these lines.  
Spectra with very high resolution of a sample of M3-M8 dwarfs 
at \caii could verify this possible discontinuity or breakdown 
of the Wilson Bappu effect in M dwarfs.

The values plotted in Figure~\ref{fig: width_sptype} are the means of
multiple measurements from each star, and reveal considerable spread
in FWHM at a given \mv.  We examined the standard deviations of the
multiple measurements for individual stars to determine if vertical
spread in Figure~\ref{fig: width_sptype} could be due to temporal
changes in the stars.  The maximum standard deviation for individual
stars is $\sim$0.76 \kms, the same variance as the observed vertical
spread among the different stars.  But the typical standard deviation
of 0.35 \kms~is considerably less than the vertical spread in
Figure~\ref{fig: width_sptype}.  Thus the distribution in FWHM among
stars at a given mass is due in part to temporal variability, but the
spread is dominated by the distribution of rotation and perhaps by the
degeneracy of mass for a given $M_V$.

Close inspection of Figure~\ref{fig: width_sptype} shows that the K
lines are generally wider than the H lines.  Figure~\ref{fig:
width_correlation} shows a histogram of the difference between the
mean FWHM of the K line and that of the H line for each star.  The
distribution is displaced toward positive values by 1.5 \kms.  This
greater width of the K line undoubtedly results from the slight
difference in the radiative transfer of the H and K emission lines.

\begin{figure}[ht!]
\begin{center}
\includegraphics[angle=90,width=0.75\textwidth]{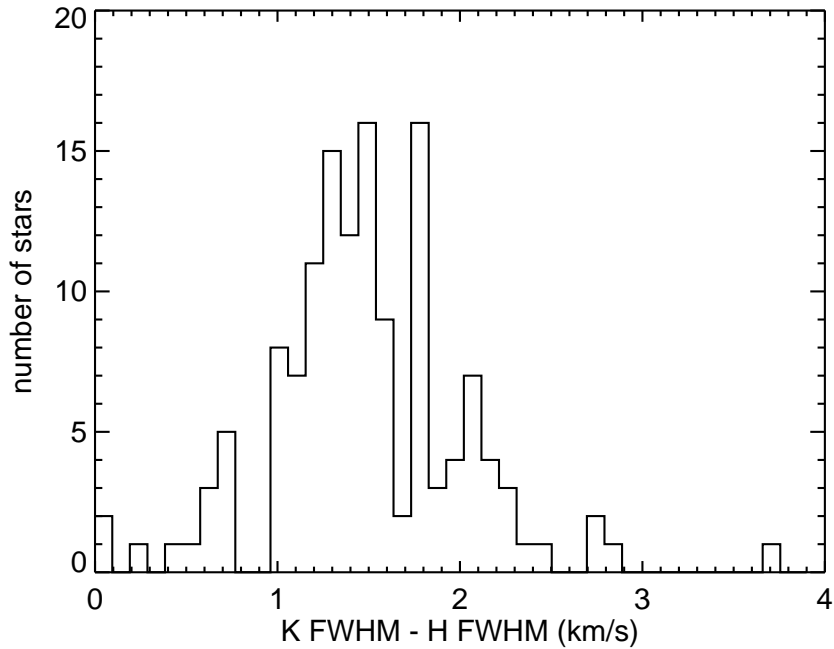}
\caption{The distribution of the difference between average FWHM of
  the H and K emission lines.  The K lines are systematically wider by
  1-2 \kms.} 
\label{fig: width_correlation}
\end{center}
\end{figure}

The \caii emission lines arise from two upper states, $^2 P_{1/2}$ and $^2
P_{3/2}$, respectively, and a common lower state, $^2S$.  The
different 2J+1 degeneracies of the two upper levels cause a slight enhancement of
the source function of the K line over the H line at each height
in the low density, upper chromosphere.  In particular, radiative 
de-excitation of the K line occurs more commonly than for
the H line. This difference causes the K line to approach the LTE
emissivity limit (the Planck function) more closely than does the H line
allowing the wings of the K line to grow faster, broadening its final profile.
However, a full description of the  differing widths of the H and K lines
requires a detailed radiative transfer model, including the pressure, temperature,
mass columns and velocity fields as a function of height in the
chromosphere, along with the partial redistribution of photon
frequencies during transitions.

\subsection{Line asymmetry}

\begin{figure}[ht!]
\begin{center}
\includegraphics[angle=90,width=0.75\textwidth]{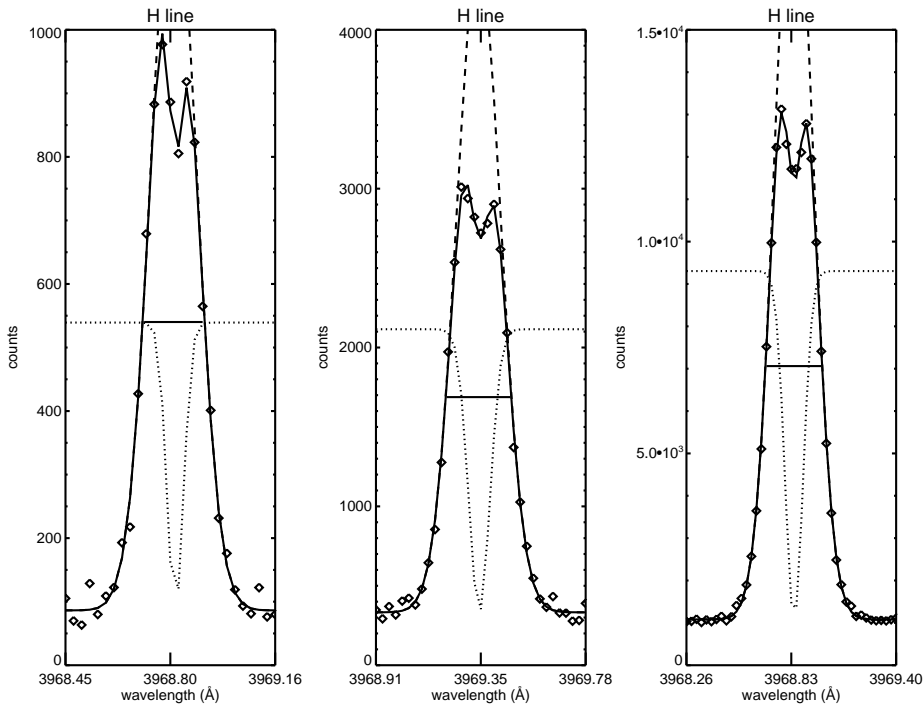}
\caption{Representative \ion{Ca}{2} H lines showing the data (dots),
  the model fits (solid line), and the component Gaussians (emission
  and absorption as dashed and dotted lines, respectively).  Line
  asymmetry from central absorption redshifts of $\Delta \lambda_c$ =
  0.4, 0.2, and 0.1 \kms~(left, middle, right) is clearly observable
  in the line profile data and model fits.  The left spectrum is 
of GJ 413, middle spectrum of GJ 424, and right spectrum of GJ 338A.} 
\label {fig: see_redshift}
\end{center}
\end{figure}

Due to the high resolution of these spectra, many of the \caii line profiles are visibly asymmetric.  Figure~\ref{fig: see_redshift} shows representative spectra from three stars with asymmetry in the \ion{Ca}{2} H line that is visible to the eye as a redshift of the central absorption (relative to the main emission).  Our V/R and $\Delta \lambda_c$ measurements permit quantitative analysis of this effect for lines with obvious asymmetry and more subtle cases.

\begin{figure}[ht!]
\begin{center}
\includegraphics[angle=90,width=0.75\textwidth]{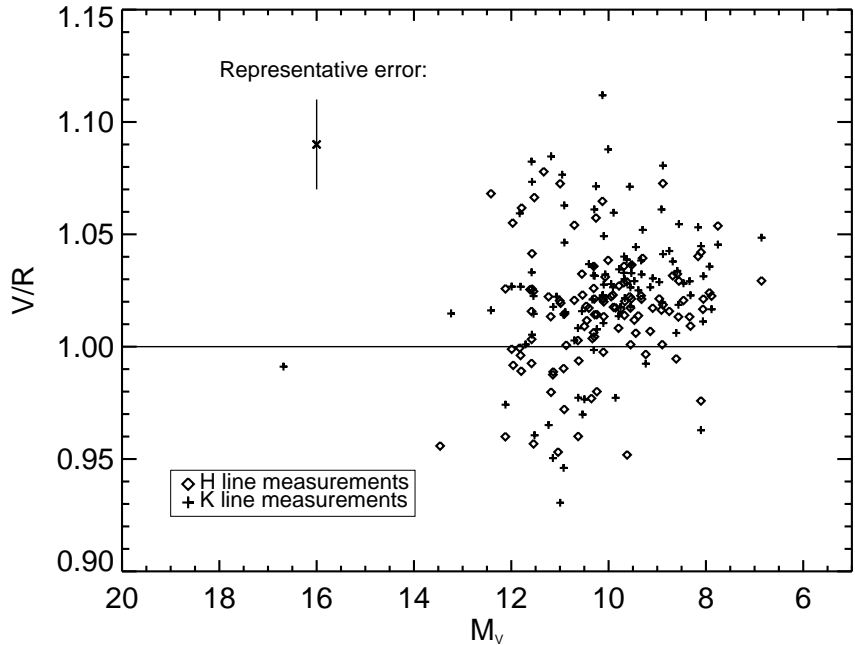}
\caption{Asymmetry measure of \caii, V/R, vs. \mv.  The solid line represents
no asymmetry, above which the domain represents lines having
redshifted central absorption.   Most K7-M5 dwarfs apparently have
redshifted absorption reversals.  Representative errors are shown for V/R.}
\label{fig: vr_sptype}
\end{center}
\end{figure}

\begin{figure}[ht!]
\begin{center}
\includegraphics[angle=90,width=0.68\textwidth]{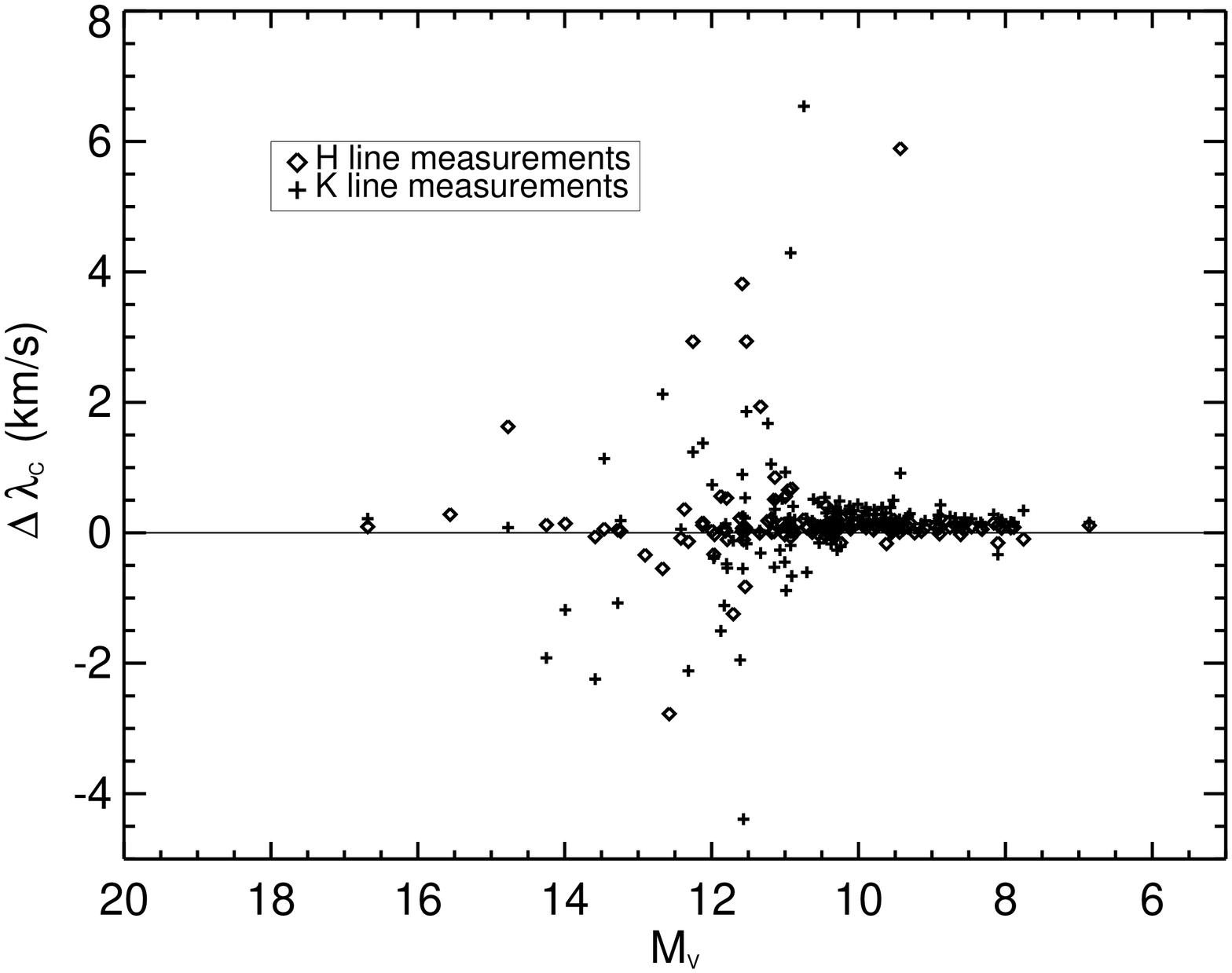}
\includegraphics[angle=90,width=0.68\textwidth]{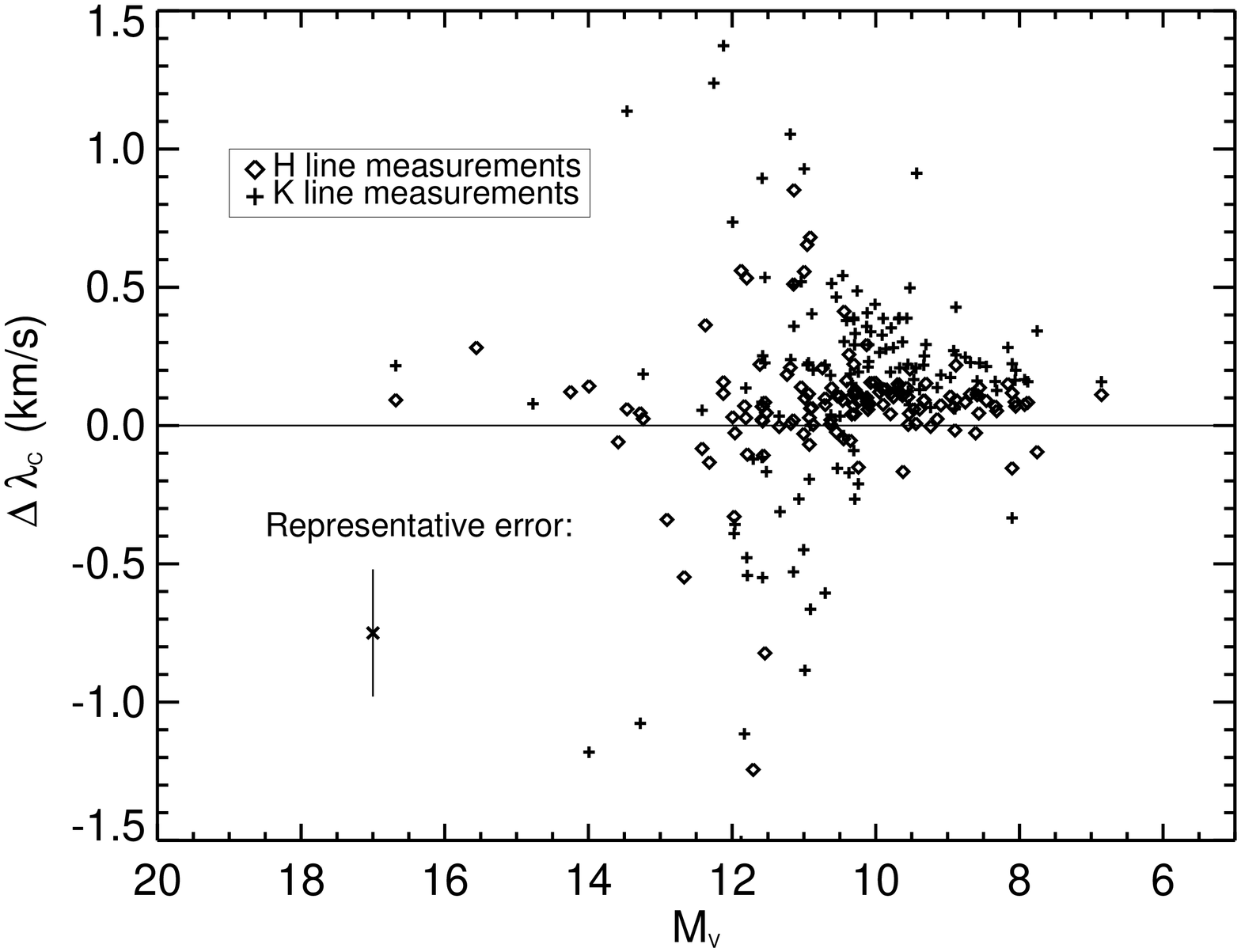}
\caption{Redshift of the absorption reversal of \caii, $\Delta
 \lambda_c$, vs. \mv.
Top: the full range of values.  Bottom:
the domain within $\pm$1.5 \kms~of symmetry.  The predominance of
points {\it above} the horizontal line shows that most K7-M5 dwarfs
have redshifted absorption reversals.
Representative errors in the top plot are about the size of the plot symbols, and are
shown in the bottom plot.}
\label{fig: shift_sptype}
\end{center}
\end{figure}

In Figures~\ref{fig: vr_sptype} and~\ref{fig: shift_sptype} we plot
V/R and $\Delta \lambda_c$ versus \mv.  The two figures show that
there is little correlation between asymmetry and absolute magnitude.
Figure~\ref{fig: shift_sptype} shows an increased spread in $\Delta
\lambda_c$ with higher \mv, i.e. for lower mass stars.  This scatter is
most likely due to the fact that these less luminous stars have
noisier lines and therefore often have lower quality fits.  Visual
inspection of the spectra with $| \Delta \lambda_c | > 2.5$ \kms~confirms 
that these spectra tend to have a lower signal-to-noise
ratio, and that the noise may be detrimentally affecting the fits.
However, these plots clearly show that most stars have V/R greater
than 1 and $\Delta \lambda_c$ greater than 0, both indicating the
presence of redshifted central absorption.

As illustrated in the bottom plot of Figure~\ref{fig: shift_sptype},
the central absorption redshifts for these stars tend to be $\sim$0.1
\kms.  Such small redshifts may seem difficult to detect, but they
correspond to 0.05 pixels, easily detectable with signal-to-noise
ratios above 20, characteristic of these spectra.  
The measured redshifts for the representative 
\ion{Ca}{2} H lines in Figure~\ref{fig: see_redshift} 
are 0.1-0.4 \kms~and clearly visible to the eye.
The atmospheres of these stars must have gas with low
source function that either has lower outward velocity or higher inward
velocity than the higher source function gas
below it.  The relative speed of 0.1 \kms~is remarkable both in its
consistency among the M dwarfs and in it small magnitude.
This speed is much slower than the escape velocity, indicating that
the gas is not falling ballistically.
Instead, it implies gas slowly decelerating upward or sinking
downward slowly, with velocities controlled by gas
pressure differences or magnetic forces.

Just as Figure~\ref{fig: width_sptype} showed possible evidence of
wider K lines than H lines, Figure~\ref{fig: shift_sptype} shows that
the K line central absorption is typically more redshifted than the H
line counterpart.  To quantify this, we plot H line asymmetry measurements
versus K line asymmetry measurements for each star in Figure~\ref{fig:
asym_corr_hk}.

\begin{figure}[ht!]
\begin{center}
\includegraphics[angle=90,width=0.63\textwidth]{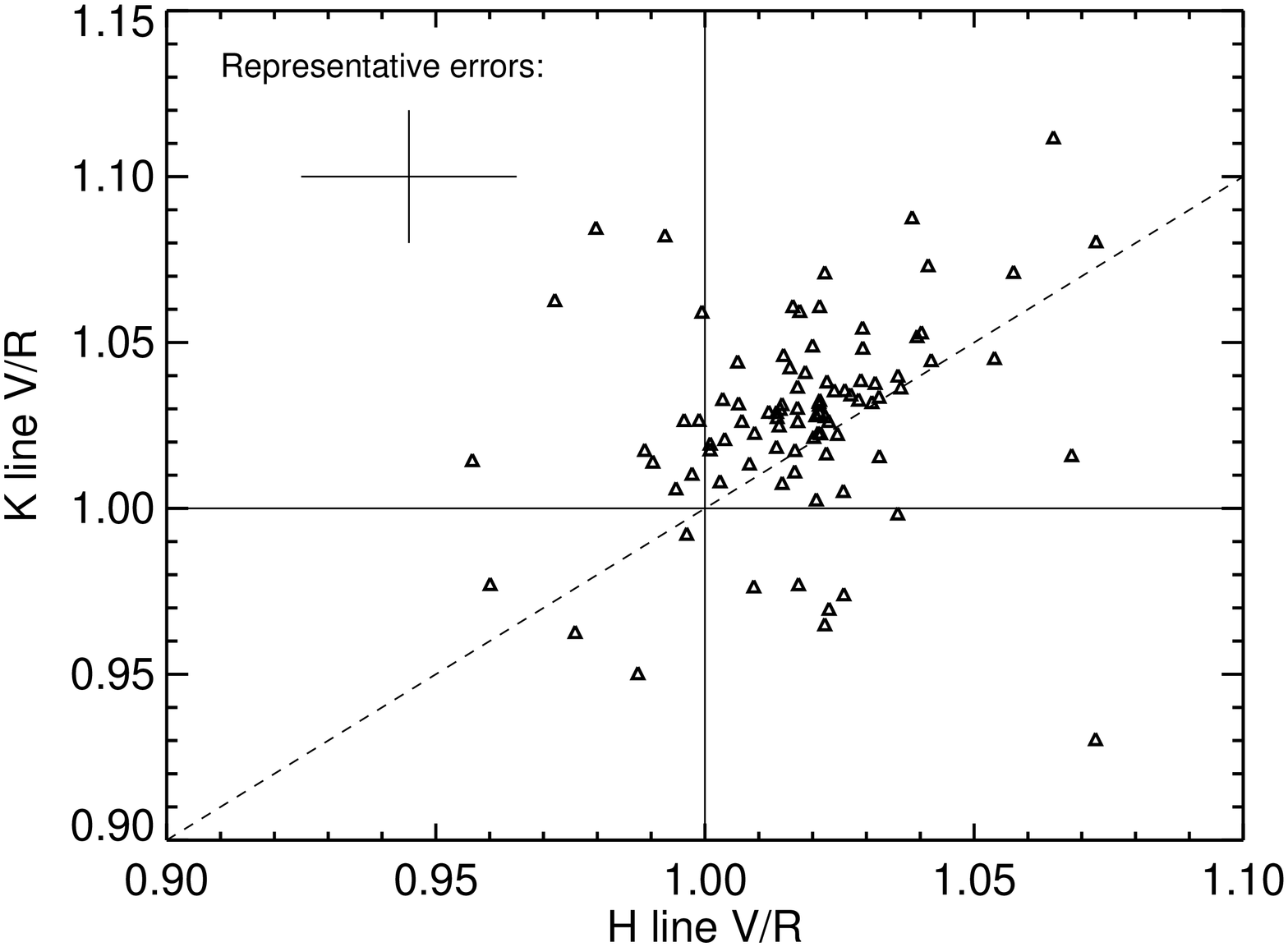}
\includegraphics[angle=90,width=0.63\textwidth]{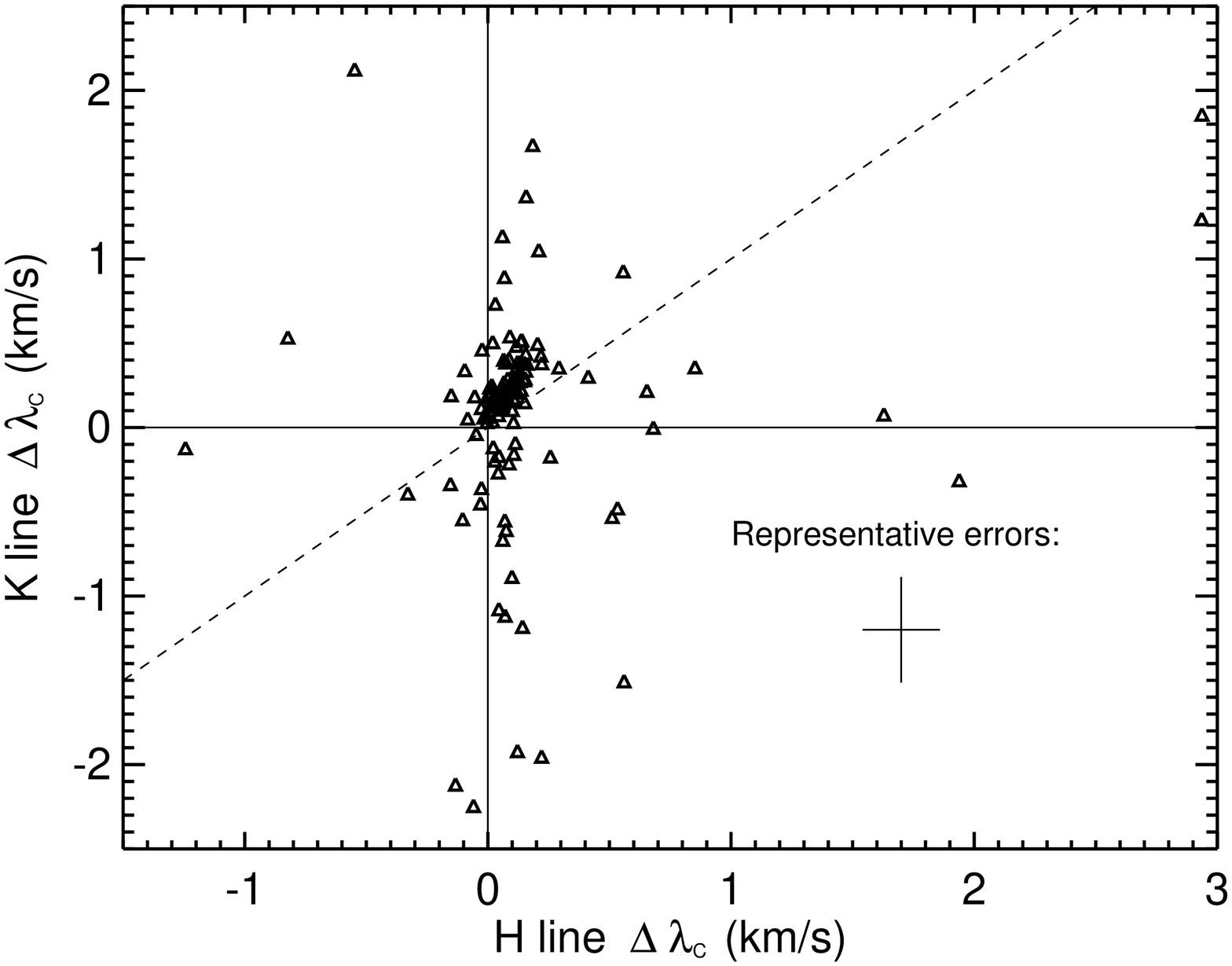}
\caption{The correlation between asymmetry in the \caii lines measured
  two ways, using V/R (top) and $\Delta \lambda_c$ (bottom).
The solid lines
separate measured redshifted absorption from blueshifted absorption.
The dashed lines show an identity relation between the two variables.  
The asymmetries in the H and K lines are clearly correlated, albeit
with scatter due to the measurement errors.
The standard deviation of the residuals to a fit of the V/R data to the identity relation is
0.024, which is roughly equal to the V/R error.  For the $\Delta \lambda_c$ data, the standard deviation is 0.7 \kms, while the measurement error is $\sim$0.23 \kms.}
\label{fig: asym_corr_hk}
\end{center}
\end{figure}

\newpage

If every star had an H line central absorption redshifted by the same
amount as its K line central absorption, we would expect to see an
identity relation in a plot of these parameters.  Although our errors
may be overestimated, visual inspection of Figure~\ref{fig:
asym_corr_hk} shows an excess of points above the identity line,
indicating a higher redshift for the absorption reversal of the K line.
This implies that the K line central absorption forms in a region with
slightly slower outward speeds (or greater inward speeds) than both
the gas where the H abosorption reversal forms and the 
gas where the H emission forms.
This result is consistent with a progression of decreasing outward
velocities, from lower to upper chromosphere.

\begin{figure}[ht!]
\begin{center}
\includegraphics[angle=90,width=0.68\textwidth]{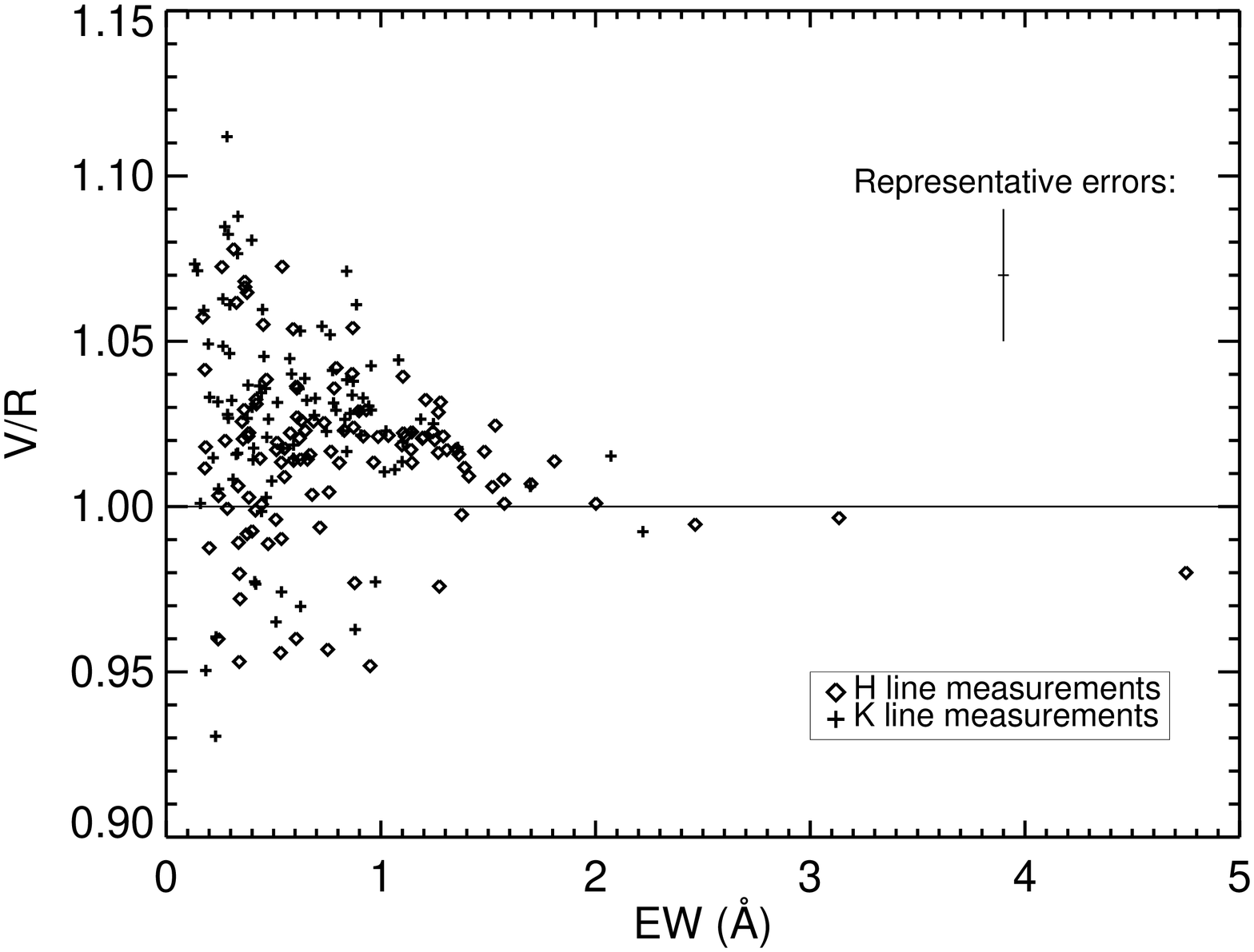}
\includegraphics[angle=90,width=0.68\textwidth]{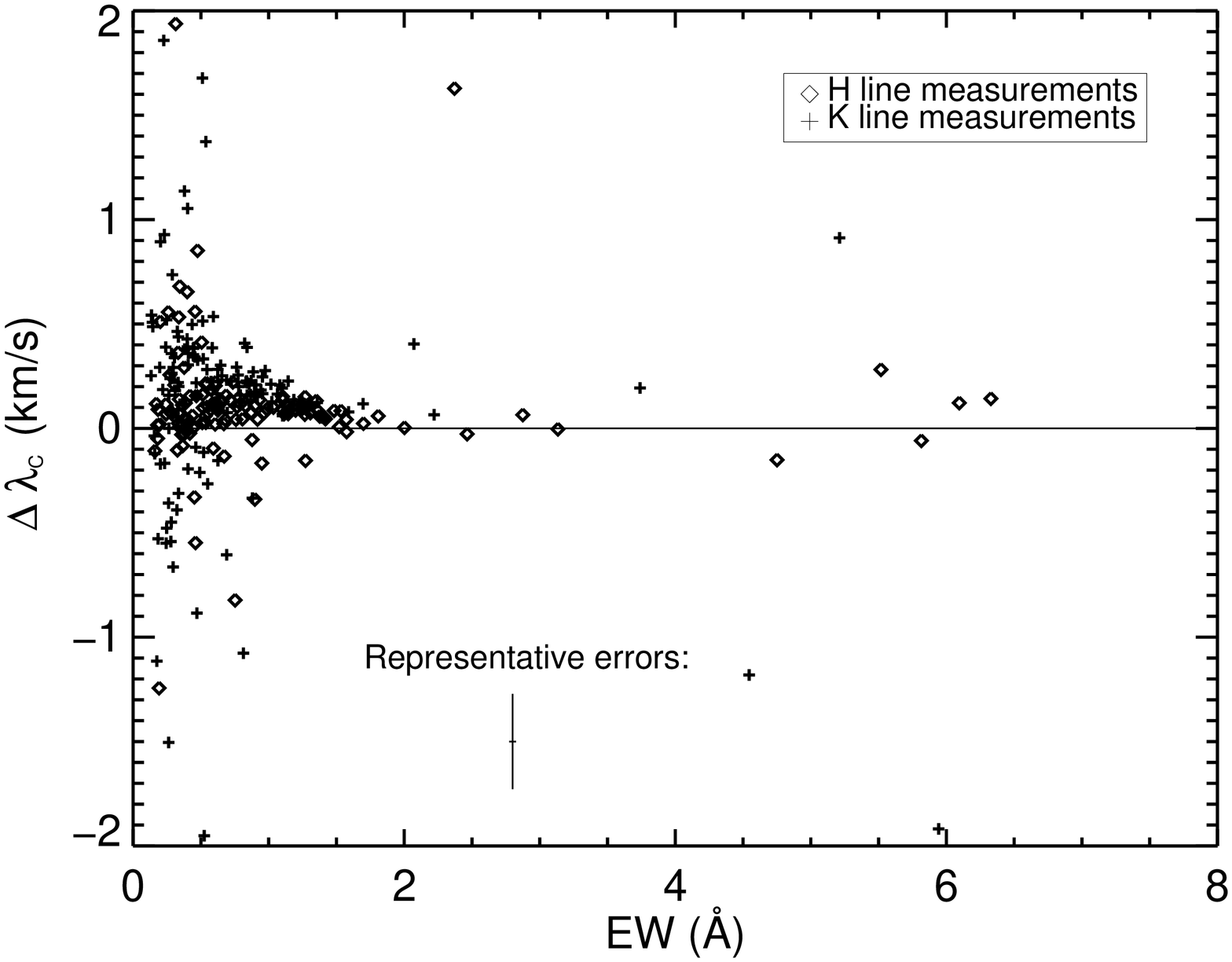}
\caption{Asymmetry measurements versus \caii EW.  The solid lines
separate emission having redshifted from blueshifted central
absorption. Most stars have a redshifted absorption reversal.  The
scatter in redshift is larger for stars with small EW, possibly a
result of the larger measurement error in asymmetry associated with
weaker emission.  Representative errors are shown in both plots.}
\label{fig: asym_ew}
\end{center}
\end{figure}

Lastly, we find an interesting possible trend between the average
asymmetry and EWs of the H and K lines.  Figure~\ref{fig: asym_ew}
shows more scatter in the asymmetry of a line at EWs less than about 1
\AA, while at EWs up to 2 \AA~the asymmetry tends to a smaller range
of slightly redshifted values.  Lines with low EWs may be weak and
therefore more likely to have noise disrupt their fits, but visual
inspection of these lines shows that this cannot account for all of
the spread seen in Figure~\ref{fig: asym_ew}.  This indicates a
possible connection between the physics governing the chromospheric
activity of a star and the velocity gradient in the chromosphere, such
that as a star becomes more active its central absorption becomes
redshifted.

\subsection{Comparison with H$\alpha$ data} \label{sec: xandha}

Here we compare the \caii equivalent widths to previously published
H$\alpha$ equivalent widths taken from \citet{Gizis2002}.  H$\alpha$
equivalent widths have long been a standard measure of chromospheric
heating in M dwarfs.  By comparing the Ca II resonance lines to
H$\alpha$ emission, the non-radiative heating and gas conditions in
the lower and upper chromosphere may be compared.

H$\alpha$ has been extensively studied with regard to chromospheric
and coronal heating in M dwarfs, indicating a peculiar behavior
\citep{Cram1979, Giampapa1982}.  With increasing magnetic fields, the
Balmer lines first become stronger {\it in absorption} before they fill in and
eventually go into emission.  In contrast, the Ca II lines increase
monotonically with increased magnetic heating \citep{Cram1979}.  Due
to this monotonic behavior, the Ca II emission strength provides a
less ambiguous diagnostic of the magnetic state of the stellar surface
than does H$\alpha$ \citep{Doyle1994}. Nonetheless, the most
magnetically active M dwarfs, notably those with strong coronal
emission, always show H$\alpha$ in emission.  In Figure~\ref{fig:
caew_haew} we plot the H$\alpha$ equivalent width against the \caii
equivalent width for our sample of M dwarfs.  The H$\alpha$ equivalent
width remains negative (absorption) near a value of about -0.2 \AA~for
a wide range range of Ca II EWs less than 2 \AA.  The H$\alpha$
absorption varies little with increasing \caii equivalent width up to
2 \AA.

\begin{figure}[ht!]
\begin{center}
\includegraphics[angle=90,width=0.7\textwidth]{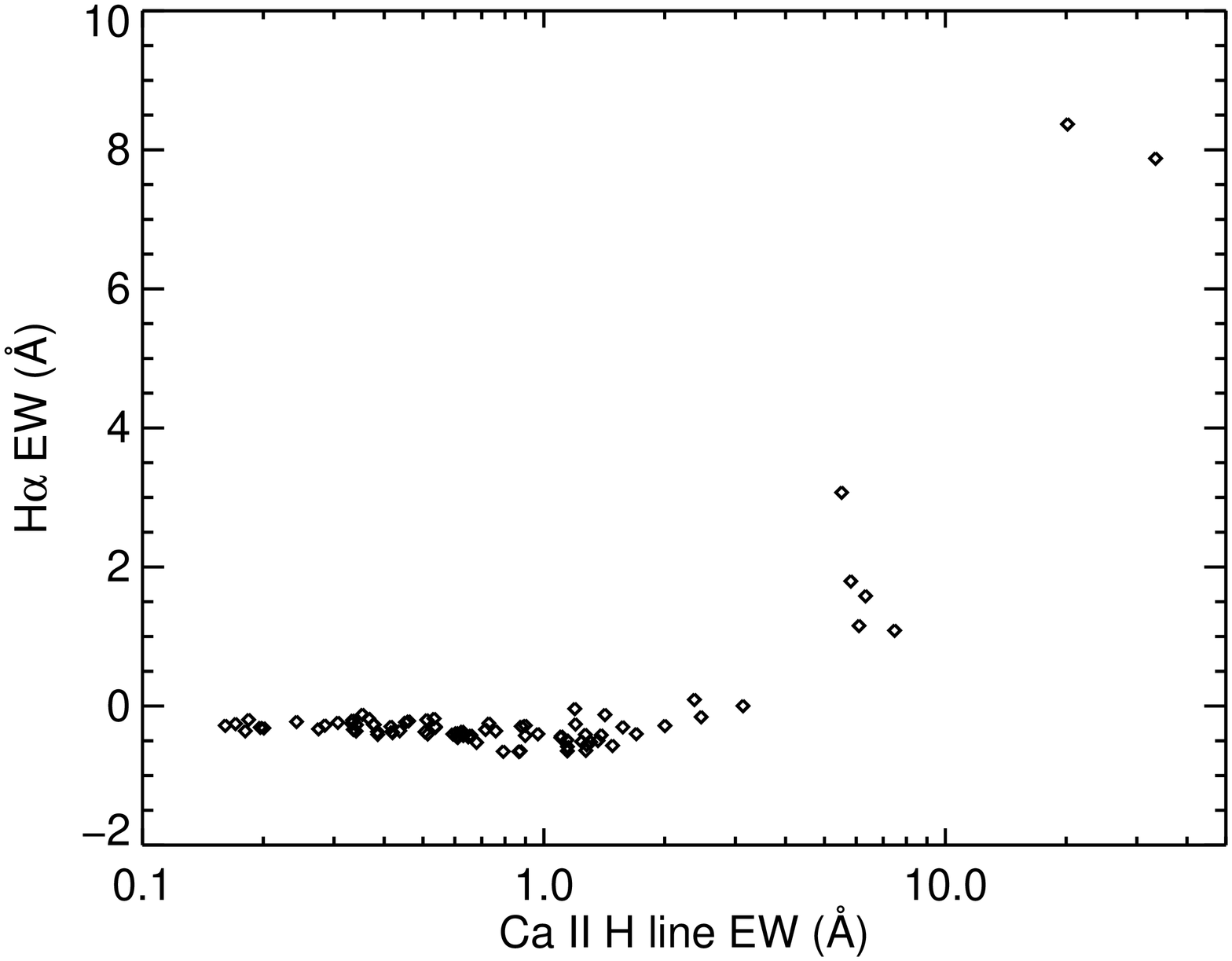}
\caption{Ca II H line EWs versus previously published H$\alpha$ EWs
for the same stars.  The two emission lines are correlated only for
large \caii emission above EW = 3 \AA.  This indicates that a magnetic 
threshold exists for H$\alpha$ while \caii remains a useful
chromospheric diagnostic at all activity levels.   H$\alpha$ EW values were taken from
\citet{Gizis2002}.  Ca II EW errors are 0.025 \AA.} 
\label{fig: caew_haew}
\end{center}
\end{figure}

However, for \caii equivalent widths above $\sim$2 \AA, H$\alpha$
abruptly goes into emission and increases monotonically with
increasing \caii .  The uncorrelated behavior between Ca II and
H$\alpha$ for stars with H$\alpha$ in absorption has been seen before
and attributed to highly segregated regions of line formation
\citep{Giampapa1989}.  \citet{Giampapa1982} found that the Ca II K
line is formed in the lower chromosphere, while the H$\alpha$ line is
formed in the upper chromosphere with a different cooling rate.  The
data presented here agree with such previously reported behavior and
reflect this segregation.  The suggestion is that the effect of
magnetic fields on the two emitting regions remains separate until a
certain level of magnetic heating is reached, at which point the lower
and upper portions of the chromosphere behave similarly.

\section{Discussion and Summary} \label{sec: discussion}

The \caii line profiles of the 143 K7-M5 dwarfs examined here revealed
various properties of their chromospheres and variations with stellar mass.
The \caii line widths and fluxes increase with increasing stellar mass and
luminosity, reminiscent of the Wilson-Bappu effect.  

The latest type stars here ($\sim$M4-M5) have \caii emission line
widths no more than 7.1 \kms~(FWHM), and possibly smaller as limited
by our spectral resolution of 5.5 \kms.  In contrast, the higher mass
M dwarfs have \caii emission line widths of 30 \kms.  Thus, the \caii
line widths vary with stellar mass by a factor of $\sim$4 (from 7.1 to
$\sim$30 \kms) over the spectral type range from M5-K7, corresponding
to a range of stellar mass from 0.3-0.55 \msun.  It is remarkable that
the line widths vary roughly as the square of stellar mass.  The
broadening seen in the \caii lines is caused either by curve of growth
effects in the line wings as the core approaches optically depth
unity, or by broadening due to velocity fields \citep{Eason1992}.
Apparently the structure of chromospheres changes significantly over
this relatively small range in stellar mass.

The \ion{Ca}{2} K line is consistently wider than the H line in all
K7-M5 dwarfs, and its central absorption is consistently more redshifted
(relative to the chromospheric emission).  While the unequal
degeneracies in the upper levels producing these lines cause somewhat
differing optical depths, these data emphasize that the H and K lines
are produced in slightly different regions of the chromosphere and are
therefore governed by slightly different line transfer and kinematics.

Most of the K7-M5 stars exhibited redshifted central ``absorption''
reversals superimposed on the emission core, with an typical shift of
0.1 \kms.  As the absorption occurs at line center, that part of the
line profile must form higher in the atmosphere near optical depth
unity at line center.  The redshifted ``absorption'' indicates that
this higher material has a lower source function and that it is moving
outward more slowly (or inward more quickly) than the lower material
that has a higher source function.

The redshifted reversal can be explained with a simple heuristic
model.  Prominence-like or micro-flare activity results in chromospheric gas
being magnetically heated and thrust outward.  The outward moving,
dense gas produces the \ion{Ca}{2} emission while the material above
it has expanded and cooled, and is moving more slowly.  That upper
material has lower density (and perhaps slightly lower temperature),
yielding a lower source function.  In such a picture, the gas
decelerates after the initial outward kick.  However, the observed 0.1
\kms~is much slower than free-fall, ballistic speeds.  Thus, the
modest velocity difference of 0.1 \kms~implies a fluid velocity
gradient that is governed by magnetic fields or small gas pressure
differences.  These small and steady velocity gradients may be compared
to the more violent kinematics that occur in major flares in M dwarfs 
\citep{Hawley2003}.  Full NTLE models with kinematics will be constructed by
Lucianne Walkowicz, in progress.

Stars with lower \caii equivalent widths exhibited a
larger spread in the redshift of the absorption component.  Between
EWs of $\sim$1--2 \AA~the redshift was confined to a small range near
0.1 \kms; stars with weaker emission exhibited a wider range of
redshifts.  Thus there appears to be some relationship between the
physical effects that govern the EW and the velocity fields.

The standard deviations from the mean values measured for both EWs and
FWHMs were greater than our measurement errors, confirming that
temporal changes do occur for these stars.  Unfortunately, the
degeneracy between \mv~and stellar mass produces spread in our plots
that cannot be distinguished from spread due to temporal variability
or the influence of individual stellar parameters (e.g. age, rotation
rate, metallicity) on chromospheric heating.

As previously found, there is clear evidence here of segregation
between the region of the chromosphere that produces H$\alpha$ and
that which produces \caii for stars with H$\alpha$ in absorption.
However, at Ca II EWs greater than about 2 \AA~we find a strong
correlation between \caii and H$\alpha$.  Apparently the upper and
lower regions of chromospheres in M dwarfs behave independently until
some magnetic activity threshold is reached, at which point they
become governed by the same physical mechanism.

The results in this paper offer empirical suggestions of structural
and kinematic attributes of M dwarf chromospheres that depend on
stellar mass and vary with height.  Models will be needed that include
the non-LTE, partial redistribution line transfer of the \caii lines
in atmospheres that contain velocity fields, to thoroughly explain the
observed properties of chromospheric line profiles.  In addition,
stellar mass estimates are required to study the relation between
chromospheric activity and mass.  Finally, contemporaneous high
resolution optical and UV spectra along with X-ray measurements would
help constrain the relationship between chromospheres and coronae in M
dwarfs.

\acknowledgements

We thank R. Paul Butler, Steve Vogt, and Jason Wright for obtaining
most of the spectra used here.  We thank Gibor Basri and Lucianne
Walkowicz for valuable conversations.  We appreciate many valuable
comments provided by the anonymous referee, making the paper much
richer.  We acknowledge support
from NASA grants NCC5-96 and NCC5-511 as well as NSF grant
HRD-9706268.  We wish to especially thank those of Hawaiian ancestry
on whose sacred mountain of Mauna Kea we are privileged to be guests.
Without their generous hospitality, the Keck observations presented
herein would not have been possible.

\begin{deluxetable}{lllcrrccc}
\tablewidth{0pt}
\tablenum{1}
\tablecaption{Observed stars}
\tablehead{
\colhead{GJ} & \colhead{HIP} & \colhead{Other} & \colhead{Spectral} & \colhead{$M_V$} & \colhead{B-V} & \colhead{Number} & \colhead{Begin} & \colhead{End} \\
\colhead{} 	& \colhead{} & \colhead{} 	& \colhead{type}	& \colhead{} 		& \colhead{} 	& \colhead{of obs.} & \colhead{(JD-2440000)} & \colhead{(JD-2440000)}
}
\startdata
  	& 428 &  			& M1 & 9.67 & 1.48 & 21 & 11757.10 & 12989.71 \\
1	& 439 & HD 225213 	& M1.5 & 10.37 & 1.46 & 3 & 12600.79 & 12899.00 \\
14	& 1368 & 			& K7 & 8.06 & 1.42 & 10 & 12097.03 & 12924.89 \\
15	& 1475 & HD 1326	& M1.5 & 10.31 & 1.56 & 25 & 10461.77 & 12924.89 \\
15B	& 	& HD 1326B	& M3.5 & 13.28 & 1.80 & 19 & 10667.00 & 13197.08 \\
1009 & 1734 & 			& M1.5 & 9.86 & 1.48 & 4 & 12600.80 & 13196.06 \\
26 	&	&			& M2.5 & 10.55 & 1.54 & 26 & 11581.71 & 12988.81 \\
27 	& 3143 &			& M0.5 & 9.62 & 1.46 & 4 & 12600.80 & 13196.08 \\
\enddata
\end{deluxetable}

\begin{deluxetable}{lllcrrrrrrrr}
\rotate 
\tablewidth{0pt}
\tablenum{2}
\tablecaption{Average H line parameters}
\tablehead{
\colhead{GJ} & \colhead{HIP} & \colhead{Other} & \colhead{\# good} & \colhead{ EW} & \colhead{ $\sigma_{EW}$} & \colhead{ FWHM} & \colhead{ $\sigma_{FWHM}$} & \colhead{ $\Delta \lambda_c$} & \colhead{ $\sigma_{\Delta \lambda_c}$} & \colhead{ V/R} & \colhead{ $\sigma_{V/R}$} \\
\colhead{} 	& \colhead{} & \colhead{} 		& \colhead{spectra \tablenotemark{a}} & \colhead{(\AA)} 		& \colhead{(\AA)} 	& \colhead{(\kms)} & \colhead{(\kms)} & \colhead{(\kms)} & \colhead{(\kms)} & \colhead{} & \colhead{}
}
\startdata
 & 428 &  &    20 &   	1.25   &      0.13   &       19.49   &      0.31   &      0.12   &     0.08   &       1.020   &     0.019  \\
1 & 439 & HD 225213 & 3 &      0.27   &     0.06   &       12.66   &      0.30 &      0.26   &      0.14   &      -- &      --\\
14 & 1368 &  &   10 &    1.48   &      0.10 &       23.82   &      0.25   &     0.09   &     0.07   &       1.017   &     0.018  \\
15 & 1475 & HD 1326 & 22 &     0.33   &     0.07   &       14.42   &      0.36   &     0.07   &     0.07   &       1.006   &     0.021  \\
15B &  & HD 1326B &    10 &   1.42   &       1.04   &       10.17   &      0.98   &     0.05   &      0.25   &      -- &      --\\
1009 & 1734 &  &   3 &    1.35   &     0.09   &       19.34   &      0.28   &      0.13   &     0.07   &       1.017   &     0.029  \\
26 &  &  &   25 &   0.42   &     0.09   &       14.20   &      0.51   &    -0.02   &      0.24   &       1.032   &     0.034  \\
27 & 3143 &  &  3 &    0.95   &      0.15   &       20.01   &      0.53   &     -0.17   &      0.23   &      0.952   &     0.096  \\
\enddata
\tablenotetext{a}{Occasionally our automated process could not fit the line(s) in a spectrum, or the fit was of bad quality and had to be rejected.  This column lists the number of spectra in which the H line was successfully fit for each star.}
\end{deluxetable}

\begin{deluxetable}{lllcrrrrrrrr}
\rotate
\tablewidth{0pt} 
\tablenum{3}
\tablecaption{Average K line parameters}
\tablehead{
\colhead{GJ} & \colhead{HIP} & \colhead{Other} & \colhead{\# good} & \colhead{ EW} & \colhead{ $\sigma_{EW}$} & \colhead{ FWHM} & \colhead{ $\sigma_{FWHM}$} & \colhead{ $\Delta \lambda_c$} & \colhead{ $\sigma_{\Delta \lambda_c}$} & \colhead{ V/R} & \colhead{ $\sigma_{V/R}$} \\
\colhead{} 	& \colhead{} & \colhead{} 		& \colhead{spectra \tablenotemark{a}} & \colhead{(\AA)} 		& \colhead{(\AA)} 	& \colhead{(\kms)} & \colhead{(\kms)} & \colhead{(\kms)} & \colhead{(\kms)} & \colhead{} & \colhead{}
}
\startdata
  & 428 &  &    20 &  0.91   &      0.10 &       20.83   &      0.58   &      0.21   &      0.21   &       1.022   &     0.031  \\
1 & 439 & HD 225213 & 3 &     0.20 &     0.04   &       13.84   &     0.01   &     -0.17   &      0.25   &      -- &      --\\
14 & 1368 &  &    10 &   1.07   &     0.07   &       25.71   &      0.47   &      0.16   &      0.16   &       1.011   &     0.019  \\
15 & 1475 & HD 1326 &  24 &    0.24   &     0.04   &       15.42   &      0.45   &      0.39   &      0.35   &       1.032   &      --\\
15B &  & HD 1326B &    8 &  0.81   &      0.54   &       10.04   &      0.86   &      -1.07   &       1.45   &      -- &      --\\
1009 & 1734 &  &   4 &   0.97   &     0.07   &       21.04   &      0.44   &      0.27   &      0.49   &      0.977   &    0.005  \\
26 &  &  &   21 &   0.33   &     0.06   &       15.33   &      0.77   &      0.46   &       1.17   &       1.016   &     0.016  \\
27 & 3143 &  & 0 &     -- &      -- &      --   &      -- &      -- &      -- &      -- &      --\\
\enddata
 \tablenotetext{a}{Occasionally our automated process could not fit the line(s) in a spectrum, or the fit was of bad quality and had to be rejected.  This column lists the number of spectra in which the K line was successfully fit for each star.}
\end{deluxetable}

\end{document}